\documentclass[twocolumn,showpacs,preprintnumbers,amsmath,amssymb]{revtex4}

\usepackage{graphicx}
\usepackage{dcolumn}
\usepackage{bm}

\begin{document}


\title{Faraday spectroscopy of atoms confined in a dark optical trap}

\author{Matthew L. Terraciano}
\author{Mark Bashkansky}%
\author{Fredrik K. Fatemi}%
\email{ffatemi@ccs.nrl.navy.mil}

\affiliation{Naval Research Laboratory, 4555 Overlook Ave. S.W.,
Washington, DC 20375}

\date{\today}

\begin{abstract}
We demonstrate Faraday spectroscopy with high duty cycle and
sampling rate using atoms confined to a blue-detuned optical trap.
Our trap consists of a crossed pair of high-charge-number hollow
laser beams, which forms a dark, box-like potential. We have used
this to measure transient magnetic fields in a 500$\mu$m-diameter
spot over a 400 ms time window with nearly unit duty cycle at a 500
Hz sampling rate. We use these measurements to quantify and
compensate time-varying magnetic fields to $\approx$10 nT per time
sample.
\end{abstract}

\maketitle

\section{\label{sec:level1}Introduction}

\noindent A spin-polarized atom sample is strongly birefringent for
near-resonant light.  This magneto-optic polarization rotation can
be used for sensitive alkali-vapor magnetometry~\cite{budker2002},
and has been the subject of several recent studies in a variety of
cold atom
samples~\cite{Isayama1999,Labeyrie2001,Zeilinger2001,jessen2003,geremia}.
When applied to localized cold atom ensembles, the result can be
sensitive magnetometry with linear spatial resolution of a few tens
of microns~\cite{vengalattore}.  These magnetic microscopes could be
of use for imaging fields near a variety of surfaces, including
integrated circuits~\cite{chatraphorn, wildermuth} and atom chips
designed for cold atom interferometry~\cite{anderson}. At a more
fundamental level, Faraday spectroscopy has also been considered for
searches of atomic electric dipole moments (EDM)~\cite{romalis} and
for nondestructive quantum state estimation and
preparation~\cite{jessen2004, chaudhury, geremia}. Such measurements
benefit from large atom numbers, long interrogation times, and
``field-free'' confinement, \emph{i.e.} confinement in which the
trapping potential minimally perturbs the measurement.

A simple way to achieve field-free conditions for a cold atom sample
is to release the atoms from a trap and probe them during freefall.
A drawback of this is that the maximum interrogation time is limited
to a few tens of milliseconds as the atom cloud falls away from the
interaction region.  Isayama \textit{et. al.}~\cite{Isayama1999}
reported a Faraday signal from atoms in freefall with a 1/$e$ decay
time of 11 ms. This limitation has been overcome by confining the
atoms to the antinodes of a red-detuned optical lattice in which one
of the lattice beams also serves as a probe beam~\cite{jessen2003}.
When the atoms were held in the intensity nodes of a blue-detuned
lattice, dark-field confinement was achieved, although the signal
was reduced because the interaction with the probe was
correspondingly diminished.

By confining atoms in a blue-detuned trap, however, it is possible
to achieve the simultaneous conditions of long interrogation time,
low-field confinement, and large atom-number~\cite{ozeri, friedman,
kaplan, kulin}. Blue-detuned traps produce lower light shifts and
photon scattering rates than red-detuned traps, enabling deep, large
volume traps with low power requirements. Although these traps have
been proposed for use in magnetometry~\cite{budker2002, Isayama1999}
and EDM searches~\cite{romalis}, to the best of our knowledge, no
experimental demonstrations have been performed.

In this paper, we report the use of dark optical traps to confine
atoms in a submillimeter, box-like volume for dynamic magnetometry
using Faraday spectroscopy. The traps are formed from crossed,
high-charge-number hollow laser beams~\cite{fatemi_ao, FatemiAOM}.
By repetitively spin-polarizing the confined sample, we extend the
measurement time from only a few milliseconds to $\approx$400 ms in
a single loading cycle with up to 1 kHz sampling rate. We
demonstrate the technique by measuring and compensating ambient
time-varying magnetic fields, such as those arising from eddy
currents and the AC power line.  We also show that nonlinear spin
dynamics due to the probe beam~\cite{jessen2004} are preserved in
these traps. The increase in duty cycle demonstrated here is
promising for both magnetometry and for efficient quantum state
preparation based on these nonlinear dynamics.

\begin{figure}[htb]
\centerline{\includegraphics[width=8cm]{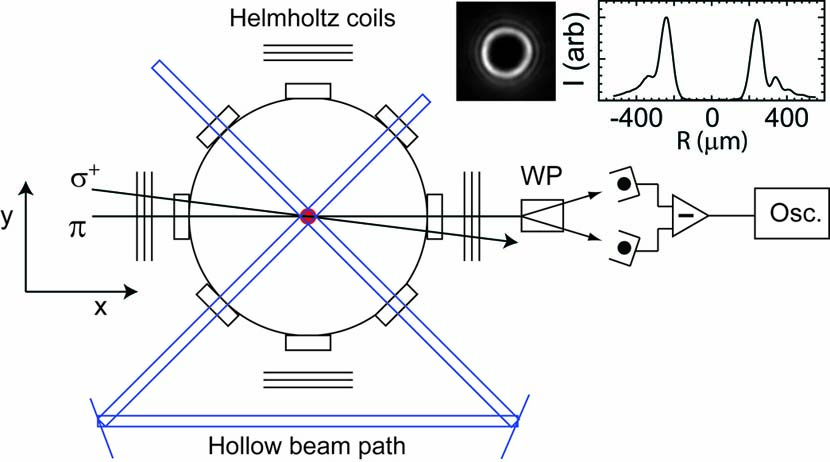}}
\caption{\label{fig:setup}(Color online) Layout of experiment.
Crossed hollow beams confine atoms to a 0.48mm diameter spot. Relay
lenses for the hollow beams are not shown. Faraday pump and probe
beams propagate along the x-axis to a balanced polarimeter. WP:
Wollaston prism. Osc:  Oscilloscope. Image of beam and cross
sectional profile are shown.}
\end{figure}

Figure~\ref{fig:setup} shows the schematic layout of the experiment.
The hollow beam is relayed to intersect itself by an $8f$ imaging
relay, as described in Ref.~\cite{FatemiAOM}.  Helmholtz coils on
all three axes control the magnetic field.  The hollow beams for our
trap are formed by modifying the wavefront phase of a Gaussian beam
with a reflective spatial light modulator (SLM). SLMs have found
increasing value in cold atom manipulation experiments because of
their ability to control trap parameters in a programmable manner
and to produce traps with nontrivial intensity
profiles~\cite{olson2007, FatemiAOM, mcgloin2003, pasienski2008,
chatt}. The applied phase for the hollow beams used here has a
profile $\Psi(\rho, \phi) = n\phi + f\lambda/(\pi\rho^2)$, where
$\rho$ and $\phi$ are cylindrical coordinates and $n$ is an integer.
The second term is a lens function of focal length $f\approx200$mm
to focus the beam of wavelength $\lambda$ onto the atom sample.  For
high charge number beams ($n\geq4$), we usually operate the trap a
few centimeters away from the focal plane, where aberrations are
reduced and the peak intensity is maximum~\cite{fatemi_ao}.

The light for the hollow beam is derived from a tunable extended
cavity diode laser.  It is amplified to 400 mW by a tapered
amplifier, 200 mW of which is coupled into polarization-maintaining
fiber.  Residual resonant light from amplified spontaneous emission
is filtered out by a heated vapor cell. The fiber output is
collimated to a 1/$e^2$ waist of 1.71 mm, and modified by the SLM
(Boulder Nonlinear Systems), which has $\approx$ 90\% diffraction
efficiency. The SLM has been calibrated at the pixel level to
correct for wavefront distortion intrinsic to the SLM.  An image of
the beam is shown in the inset to Fig.~\ref{fig:setup}.  Our choice
of $n=8$ is driven by the practical considerations of field-free
confinement and large trap size, but these trap parameters can be
adjusted with the SLM.

Our experiment begins with cold $^{85}$Rb atoms derived from a
magneto-optical trap (MOT).  We confine $\approx10^7$ atoms in a
$\approx500\mu$m diameter (1/$e^2$) cloud.  The atoms are further
cooled in a 10 ms long molasses stage to $\approx10\mu$K, after
which all MOT-related beams are extinguished.  The hollow beam trap
is on throughout the MOT loading, but can be switched off by an
acoustooptic modulator. The Faraday spectroscopy is performed by
similar technique as in Ref.~\cite{Isayama1999}. To perform these
measurements, a pair of laser beams is used along the $x$-axis
(Fig.~\ref{fig:setup}). The atoms are optically pumped into the
$F=3, m_F = 3$ stretched state by a 20 $\mu$s $\sigma^+$ pulse
connecting $F=3 \rightarrow F' = 3$. This beam has 1/$e^2$ waist of
6.0 mm and has a peak intensity of $\approx3I_{sat}$, where
$I_{sat}$ is 1.6 mW/cm$^2$.  This beam is retroreflected to prevent
unidirectional momentum kicks, and a small amount of repumper light
($\approx0.04I_{sat}$) is added during this pulse to keep the atoms
in the $F=3$ hyperfine ground state. When this light is
extinguished, the atoms begin precessing freely at the Larmor
precession frequency $\omega_L=g_F\mu_BB/\hbar$, where $g_F$ is the
gyromagnetic ratio, $\mu_B$ is the Bohr magneton, and $B$ is the
magnetic field.  For $^{85}$Rb, $g_F\mu_B/\hbar=466.7415$
kHz/Gauss~\cite{alexandrov}. A linearly polarized probe beam at a
detuning $\Delta_p=2\pi\times2.5$ GHz with $\simeq$20mW and 1/$e^2$
waist $\omega_p$ = 6.0mm passes through the atom cloud to a simple
polarimeter consisting of a Wollaston prism that splits the probe
beam into two orthogonal polarization states that are detected by a
balanced photodetector. For these parameters, the photon scattering
time from the probe beam is calculated to be $\tau_{p}\approx$2ms.
The MOT region is imaged onto a pinhole along the axis of the probe
beam so that only the portion of the probe that interacts with the
confined atoms reaches the detector.

The hollow beam trap prevents the atoms from falling away from the
interaction region during the probing process.  The beam has 150 mW
total power at the trap.  We use a detuning $\Delta_{t}
\approx25$GHz (=0.05 nm) above the $F=3 \rightarrow F' = 4$
transition. At the MOT, the hollow beam has a diameter of 0.48 mm,
measured between maxima, and the peak intensity is $8.2\times10^4$
mW/cm$^2$ for a trap depth $U\approx2\hbar\Gamma\approx3000E_r$,
where $E_r$ is the recoil energy for $^{85}$Rb. The gravitational
potential energy across this trap is $\hbar\Gamma/6$. For these
parameters, the peak scattering rate from the trapping beams would
be $\gamma_{t}=1/\tau_{t}\approx2\pi\times3$kHz, but is reduced from
this value by being trapped in the dark. Although we do not measure
this value, we establish an upper bound to be
$\gamma_{t}{\leq}2\pi\times200$Hz.

\begin{figure}[htb]
\centerline{\includegraphics[width=8cm]{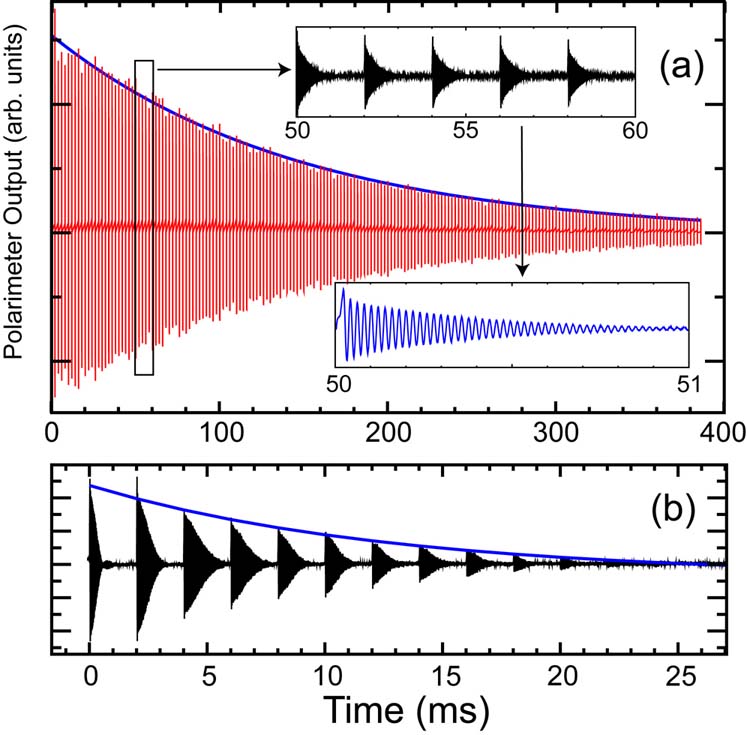}}
\caption{\label{fig:longscan} (Color online) Faraday signals for a)
trapped sample; and b) untrapped sample.  The untrapped atoms fall
away from the probe region within 20 ms, while the trapped atoms
remain with 150 ms time constant. Insets to (a) are expanded views
of the raw data.}
\end{figure}

Each optical pumping event initiates the Larmor precession.
Figure~\ref{fig:longscan}a shows 64 averages of 200 optical pumping
cycles spaced 2 ms apart in the presence of the hollow beam trap and
a bias magnetic field of $\approx100$ mG along the $z$-axis. A
single Larmor precession signal is shown in the lower inset to
Fig.~\ref{fig:longscan}a.  The envelope over all Larmor precession
signals decays with a 1/$e$ time constant of $\approx150$ ms. This
decay is due primarily to the steady heating that occurs during each
optical pumping cycle, which gradually boils atoms out of the trap.
In contrast, Fig.~\ref{fig:longscan}b shows the signals without the
hollow beam trap present.  In this case, the atoms fall completely
out of the probe beam detection window within 25 ms, with a 1/$e$
decay time of 13 ms, similar to that reported in
Ref.~\cite{Isayama1999}.  The signals in Fig.~\ref{fig:longscan} are
recorded immediately following the molasses phase of the MOT loading
cycle.  Over the first few pumping cycles, the envelope of the
individual precession signals in Fig.~\ref{fig:longscan}b changes
dramatically due to residual eddy currents in the vacuum chamber.
Holding the atoms in an optical trap allows measurements to be
performed after eddy currents have subsided, while also
substantially increasing both the measurement window and the overall
duty cycle.

For our parameters, each independent Larmor precession signal
dephases with a submillisecond 1/$e$ decay time. This dephasing
occurs from several factors, including spatial gradients and photon
scattering from the trap and probe beams.  Nonlinear Hamiltonian
terms can also shorten the decay time of the signal, as described in
Ref.~\cite{jessen2004}.  These nonlinear terms depend on the angle
between the polarization of the probe laser and the magnetic field.
When the relative angle is $\approx$ 54$^{\circ}$, the effects of
these terms are eliminated. For this work, we operated at this
relative orientation so that the dephasing occurs primarily through
photon scattering.  From Fig.~\ref{fig:longscan}, we find that the
untrapped signals decay with a 1/$e$ time of $\approx$0.7ms.  For
the samples trapped in the hollow beam, we observe a slight
reduction in the decay time to $\approx$0.5 ms. Thus we have an
upper bound for $\gamma_t\leq2\pi\times200$Hz.

\begin{figure}[htb]
\centerline{\includegraphics[width=8cm]{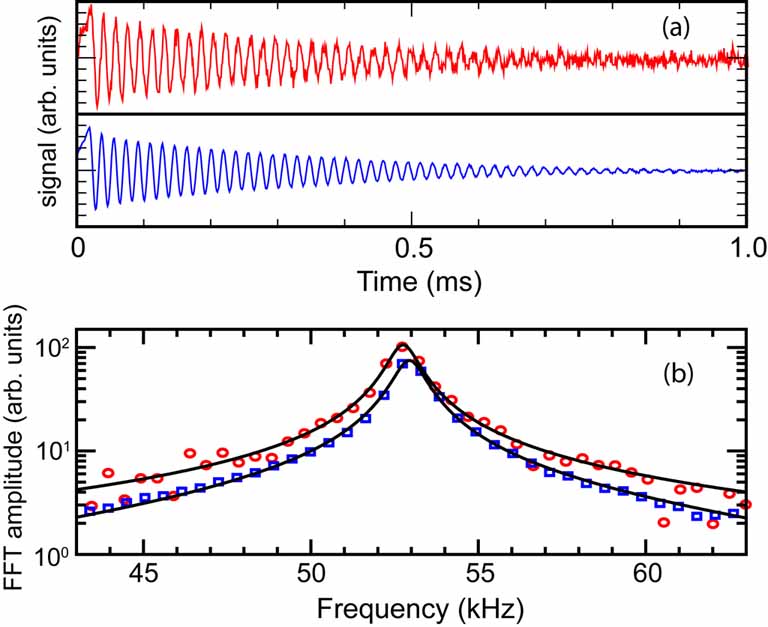}}
\caption{\label{fig:time-and-ffts} (Color online) a) Polarimeter
output for single-shot data (top) and averaged data (bottom).  b)
Fast-Fourier transform (FFT) of data in (a). Fits to a Lorentzian
profile are shown as solid lines. Single-shot (circles) and averaged
(squares) data are shown.}
\end{figure}

In a gradient-free, static magnetic field, the voltage output of the
polarimeter is an exponentially-damped sinusoid, $V(t) =$
Aexp(-$t/\tau$)sin$(2\pi\nu_Lt + \phi)$, where A is the initial
amplitude, $\tau$ is the 1/$e$ decay time,
$\omega_L=2\pi\times\nu_L$ is the Larmor frequency, and $\phi$ is a
phase.  To determine $\nu_L$, the averaged data in each 2 ms probing
window (Fig.~\ref{fig:time-and-ffts}a) are Fourier transformed
(Fig.~\ref{fig:time-and-ffts}b). We fit these transforms to a
Lorentzian, the center of which is $\nu_L$.

\begin{figure}[htb]
\centerline{\includegraphics[width=8cm]{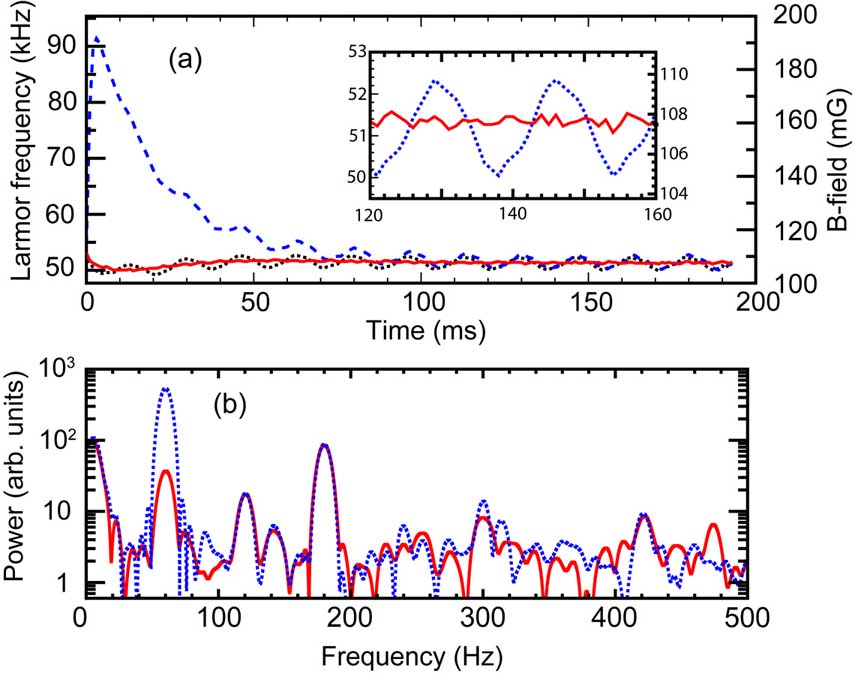}}
\caption{\label{fig:larmorfrequency}(Color online) a) Larmor
precession frequency as a function of time with various levels of
compensation. Dashed: no compensation; dotted: compensation of eddy
currents, and solid: full compensation. Inset to (a) shows magnified
view of compensation. Fluctuations in our Larmor precession are
dominated by uncompensated harmonics of the AC line.  b) Magnetic
field spectrum with no compensation (dotted), and 60Hz compensation
(solid).}
\end{figure}

In Fig.~\ref{fig:larmorfrequency}a, we plot $\nu_L(t)$ over 200
pumping cycles~(64 averages) spaced 1 ms apart (dashed line).  The
signal displays two dominant sources of time-dependence. First, the
exponential decay occurs from the metal vacuum chamber, which
develops eddy currents when the
MOT coils are extinguished. 
Due to the symmetry of our chamber, the eddy currents are along the
axis of the MOT coils ($z$).  The bias field of $\nu_L\approx50$kHz
for these measurements is also on the $z$-axis so that the eddy
current field adds linearly to the bias field. The second source of
time-dependent behavior is ambient AC magnetic fields in the room
arising from power supply transformers, power strips, etc. We note
that our experiment is triggered off the AC power line. We found
that this field is also primarily along the $z$-axis, because the
amplitude of the oscillation signal is independent of this bias
field.  An orthogonal component would add in quadrature and cause
the amplitude to vary with the bias field. Additionally, an
orthogonal, oscillating magnetic field component added in quadrature
would show up at twice its oscillation frequency.  Since the Larmor
frequencies retrieved from Faraday spectroscopy determine the scalar
magnetic field, full vector information is not acquired in a single
shot, but can be acquired through multiple
measurements~\cite{terraciano}. Some information about magnetic
field orientation can be obtained directly from the polarimeter
signal (e.g. there is no spin precession if the $B$ field is
parallel to the optical pumping axis), but that effect is outside
the scope of this work.

For many applications, control over the magnetic field is required
to sub-mG levels, especially those involving Raman transitions
between magnetically sensitive states~\cite{boyer:043405, Garreau,
terraciano, kerman}.  As a simple application of the long
measurement time capability, we demonstrate compensation of these
time varying fields.  We first compensate the effects of eddy
currents, which produce an exponentially decaying magnetic field at
the atom sample. This field decays with a 1/$e$ time of $\approx$20
ms (Fig.~\ref{fig:larmorfrequency}a). For a given MOT coil current
setting, the eddy current amplitude is constant. We produce an
opposing time-varying field flux by using a voltage-controlled
current source (Kepco ATE15-15M).  This current passes through a
20-turn Helmholtz pair of diameter 20 cm, width 2.5 cm, and
separation 11.4 cm oriented along the MOT coil axis. The appropriate
time variation is done by low-pass filtering of a step function
whose amplitude is adjusted for optimum compensation. The result is
shown in Fig.~\ref{fig:larmorfrequency}a (dotted line). Although
this source of time variation is not canceled perfectly, the field
beyond 25 ms is constant to within the 60 Hz field amplitude.

The ambient AC magnetic fields are primarily due to 60 Hz power line
sources.  By triggering our experiment from the power line, this
source of magnetic field variation is reproducible and can be
compensated. Without this triggering, the variations of a few mG
observed in Fig.~\ref{fig:larmorfrequency}a would lead to
significant shot-to-shot fluctuations of the field measurements.  We
produce an opposing field by adding a 60 Hz sinusoidally varying
current to the bias coils. The current amplitude and phase are
adjusted for optimum compensation. The result with all compensations
applied is shown in the inset to Fig.~\ref{fig:larmorfrequency}a.
The signal remains constant to within a standard deviation of 110 Hz
(230 $\mu$G). Most of this residual field is due to higher AC line
harmonics; in Fig.~\ref{fig:larmorfrequency}b, we show frequency
spectrum of the magnetic field, which clearly shows higher harmonics
at 180, 300, and 420 Hz. We suppressed the 60 Hz component by a
factor of 20. With appropriate signal processing, the field
measurements in our setup could be made in real time (with
single-shot measurements as in Fig.~\ref{fig:time-and-ffts}) and be
used as feedback control with a bandwidth determined by the
Helmholtz compensation coils.

\begin{figure}[htb]
\centerline{\includegraphics[width=8cm]{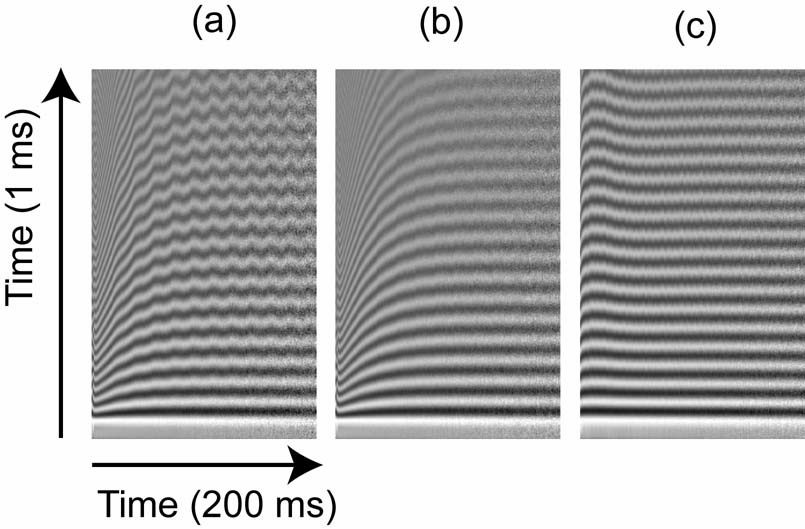}}
\caption{\label{fig:images} 2D image of data that exposes
qualitative magnetic field features without FFT processing. (a)
uncompensated, (b) 60 Hz compensation, c) full compensation.}
\end{figure}

Another way to visualize the time dependent signals, shown in
Fig.~\ref{fig:images}a, is by converting the 1D data set of
Fig.~\ref{fig:longscan} to a 2D matrix. Each successive column
contains the Larmor precession signal for subsequent triggers. This
exposes time variations in an easily identifiable way with no FFT
analysis.  We show these images for the magnetic fields with no
compensation, 60 Hz compensation and full compensation. A constant
magnetic field shows up as a series of horizontal lines whose
spacing is inversely proportional to $\omega_L$
(Fig.~\ref{fig:images}c).

\begin{figure}[htb]
\centerline{\includegraphics[width=8cm]{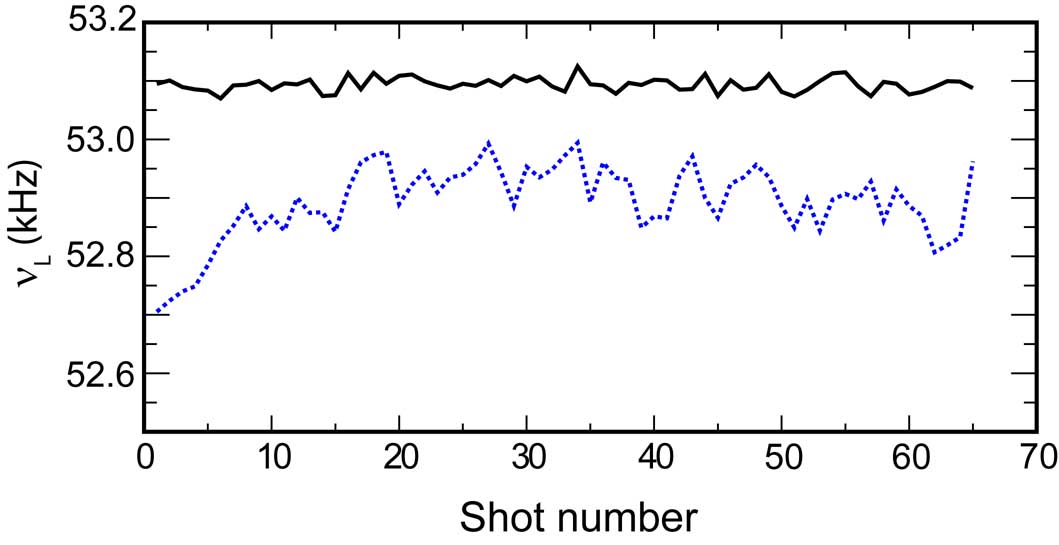}}
\caption{\label{fig:systematic}(Color online) Comparison of 64 data
runs (dashed) to a simulated signal with white noise added to have
similar SNR as the experiment (black).  For clarity, the signals are
offset from each other.}
\end{figure}

Because of the uncompensated field variations in our lab, our
measurement uncertainty is dominated by systematic errors.  To
differentiate the systematic error from the random error, we measure
the Larmor precession frequency in a 1 ms measurement window for 64
independent loading cycles at a trapping time of T = 100 ms. These
scans are recorded at a 1 Hz rate. The result is shown in
Fig.~\ref{fig:systematic}. There is a long term drift in our lab on
the order of several seconds. Our time-domain signals have a
signal-to-noise ratio (SNR) of $\approx$15
(Fig.~\ref{fig:time-and-ffts}a).  For comparison, we simulated the
expected Larmor signals for exponentially-damped sinusoids with the
same SNR that had additive white noise (Fig.~\ref{fig:systematic}).
For the experimentally measured case, we found a standard deviation,
or single-shot error, in each 1 ms optical pumping cycle of
$\delta\nu_L\approx$45Hz, or $\delta{B}\approx100\mu$G (=10 nT), and
for the simulated case, the error was $\delta\nu_L\approx$16 Hz, or
$\delta{B}\approx30\mu$G (=3 nT). This discrepancy is likely due to
other sources, such as unwanted variations in the MOT coil current.

Our hollow beam traps are initially loaded with $N\approx10^6$
atoms. The shot-noise-limited magnetic field measurement error due
to atom number is $\delta{B} \simeq
\left(\hbar/g\mu_B\right)(1/\sqrt{N\tau{T_m}})$, where $\tau$ is the
spin-coherence time and $T_m$ is the measurement
time~\cite{budker2002}.  Because we are measuring a rapidly varying
field, $T_m = 1\rm{-}2$ms, limiting $\delta{B}\approx2\mu$G (=200
pT) in each optical pumping cycle. After T = 400 ms of trapping
time, when there are only $\approx10^5$ atoms remaining, this
increases to $\approx6\mu$G (=600 pT).  Our measured values are
above the shot noise limit due to the simple photodetection circuit
we used and to incomplete optical pumping, which effectively reduces
N.

For static magnetic fields, each measurement cycle through the total
trap time can be averaged, effectively increasing $T_m$ to several
hundred milliseconds and greatly increasing the shot-noise-limited
sensitivity.  Likewise, $\tau$ can be increased by using larger
detunings for the probe and trapping beams.  These blue-detuned
traps are capable of capturing large enough atom numbers that
measurements in the low pT range or better should be possible in a
single MOT loading cycle over the entire measurement window.

\begin{figure}[htb]
\centerline{\includegraphics[width=8cm]{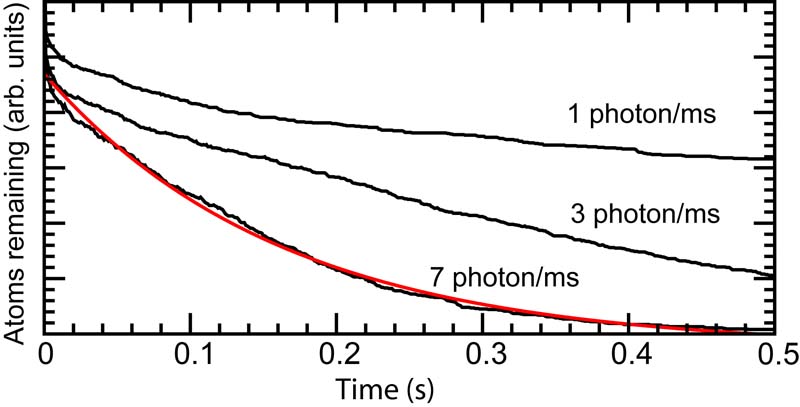}}
\caption{\label{fig:boiling}(Color online) Simulations of atom
number remaining for different degrees of optical pumping.  With no
optical pumping or probe beams (longest-lived curve) the atoms still
scattered $\approx$1 photon/ms from the trap beams. A curve fit is
shown for the case of scattering 7 photons/ms.  This time constant
of 160 ms agrees well with our experimental value of 150 ms
(Fig.~\ref{fig:longscan}).}
\end{figure}

For any trap depth, there is a trade-off between SNR and the number
of possible field measurements allowed before the signal decays. SNR
improves by increasing the number of atoms that are optically pumped
or by decreasing the pump detuning~\cite{jessen2003}, but these
approaches also boil the atoms out of the trap more quickly. For
most of our results presented here, we only weakly pumped the atoms
to reduce heating and to increase the number of optical pumping
cycles we could achieve.  In general, the dominant heating will
occur from the $20\mu$s optical pumping phase of each cycle, during
which several photons are scattered.  As a rough estimate, the
timescale for signal decay should be on the order of the time
required for the average atom energy to equal the trap depth. This
will occur after a time $T_{boil} = U/(\gamma_{tot}E_r)$, where U is
the trap depth, $\gamma_{tot}$ is the total scattering rate
(including probe, trap, and optical pumping beams), and $E_r$ is the
recoil energy. For our trap of $U~\approx3000E_r$, and assuming
$\approx10$ scattered photons every 2ms optical pumping cycle, this
gives $\approx300$ pumping cycles before the atoms are boiled away.

To examine this boiling process more accurately, we perform Monte
Carlo simulations of the atom dynamics within our trap for different
total scattering rates.  Within each time step, the atom's momentum
is changed with a probability determined by the local scattering
rate for the probe and trap beams, as calculated by the
Kramers-Heisenberg formula~\cite{MillerPRA1993}.  We performed these
simulations for different optical pumping rates.  In
Fig.~\ref{fig:boiling} we plot the number of atoms remaining as a
function of time for varying degrees of scattering rates.  For the
case of 7 photons scattered every millisecond, we find an
exponential decay of $\approx$160 ms, which is close to our observed
value of $T_{boil} = 150$ms. By turning off the probe and optical
pumping beams in the simulation, we find that the trap beam
scattering rate is $\gamma_t\approx2\pi\times100$Hz which agrees
with our upper bound of $\gamma_t\leq2\pi\times200$Hz from the
Faraday decay time.

Within our measurement error, we observed no effect of the trapping
light on the Larmor frequency.  Optically-induced Zeeman shifts that
occur with elliptically polarized light~\cite{vengalattore, romalis,
park2002} should be small, because the trap beam polarizations are
linear and because of the low field confinement. Furthermore, any
vector light shifts from the trap beams, confined to the $x-y$
plane, would add in quadrature to our applied magnetic field along
$z$, reducing the effect on $\omega_L$~\cite{romalis}. We are
currently studying the effects of trap geometry on the Larmor
precession signals.

\begin{figure}[htb]
\centerline{\includegraphics[width=8cm]{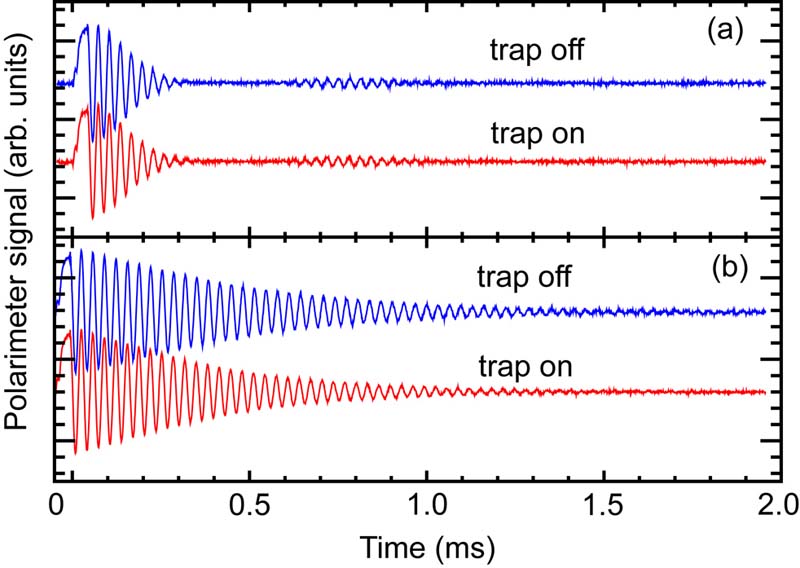}}
\caption{\label{fig:trap_on_trap_off}(Color online) Larmor
precession signals with and without the trap light at two different
relative orientations of the laser polarization with respect to the
magnetic field. These were done at a) 0 degrees (revivals maximized)
and b) 54 degrees (revivals suppressed).}
\end{figure}

The dominant source of nonlinear effects due to the laser fields is
the probe light.  As discussed in Ref.~\cite{jessen2004}, the tensor
component to the light shift adds a nonlinear term to the spin
Hamiltonian, whose magnitude is dependent on the angle between the
laser polarization and the magnetic field. This Hamiltonian plays an
important role in studies of quantum chaos and is useful for both
nondestructive quantum state preparation and
measurement~\cite{jessen2004}.  For sufficiently large magnetic
fields, the nonlinearity vanishes when the relative angle is
$\theta$=arctan$(\sqrt{2})\approx54^{\circ}$, but is maximized for
$\theta=0$. We have verified that these nonlinear spin dynamics,
which manifest themselves as revivals of the Faraday oscillation
signal, can still be observed in these hollow beam traps. In
Fig.~\ref{fig:trap_on_trap_off}, we show the Larmor precession
signals for $\theta=0$ and $\theta=54^{\circ}$ both with the trap on
continuously and with the trap switched off immediately prior to the
optical pumping pulse.  Thus the high duty cycle of the technique
presented here may be of use for rapidly testing quantum state
preparation procedures employing this nonlinearity.

We have demonstrated Faraday spectroscopy with high repetition rate,
long measurement time, and submillimeter spatial resolution in a
dark hollow beam optical trap. We used high-charge-number hollow
laser beams to provide box-like confinement with near resonant light
and low laser power. These traps can be sufficiently deep that
several hundred Faraday measurements are possible before atoms are
heated over the confining potential. We demonstrated a continuous
magnetic field measurement over a period of 400 ms which enabled us
to measure and compensate for time-varying magnetic fields. This
work was funded by the Office of Naval Research and by the Defense
Advanced Research Projects Agency.

\bigskip


\end{document}